\documentclass{iopart}
\eqnobysec
\usepackage{iopams}  
\usepackage{revsymb}  
\usepackage{graphicx,color}  
\usepackage{bm}  
\usepackage{cite}  
\newcommand{\ket}[1]{|{#1}\rangle}
\newcommand{\bra}[1]{\langle{#1}|}

\newcommand{\ketbras}[3]{\ket{#1}_{#3}\hspace*{-0.2mm}\bra{#2}}

\renewcommand{\labelenumi}{\arabic{enumi}.}

\begin{document}
\title[Entanglement by repeated on- and off-resonant scattering of ancilla qubits]{Efficient generation of a maximally entangled state by repeated on- and off-resonant scattering of ancilla qubits}
\author{Kazuya Yuasa$^1$, Daniel Burgarth$^2$, Vittorio Giovannetti$^3$ and Hiromichi Nakazato$^4$}
\address{$^1$ Waseda Institute for Advanced Study, Waseda University, Tokyo 169-8050, Japan}
\address{$^2$ IMS and QOLS, Imperial College, London SW7 2BK, UK}
\address{$^3$ NEST-CNR-INFM \& Scuola Normale Superiore, piazza dei Cavalieri 7, I-56126 Pisa, Italy}
\address{$^4$ Department of Physics, Waseda University, Tokyo 169-8555, Japan}
\ead{\mailto{yuasa@aoni.waseda.jp}}
\begin{abstract}
A scheme for preparing two fixed non-interacting qubits in a maximally entangled state is presented.
By repeating on- and off-resonant scattering of ancilla qubits, the state of the target qubits is driven from an arbitrary initial state into the singlet state with probability 1 (perfect efficiency).
Neither the preparation nor the post-selection of the ancilla spin state is required.
The convergence from an arbitrary input state to the unique fixed point (mixing property) is proved rigorously, and its robustness is investigated, by scrutinizing the effects of imperfections in the incident wave of the ancilla, such as mistuning to a resonant momentum, imperfect monochromatization, and fluctuation of the incident momentum, as well as detector efficiency.
\end{abstract}
\submitto{New J. Phys.}

\section{Introduction}
\label{sec:Introduction}
How to prepare a quantum state?
It is a nontrivial and important problem to be tackled.
In fact, various interesting and peculiar phenomena are predicted on the basis of highly nonclassical states, and entanglement plays key roles in quantum information protocols \cite{ref:QuantumInfo}. 
They all rely on the generation of nontrivial states and are not realized without establishing the strategies for the preparation of such quantum states.

Generally speaking, we try to drive a quantum system to a specific state by a series of operations, e.g., applications of external fields to transform its state, measurements to project it onto a particular configuration, and so on.
A generic mechanism was found to extract a pure quantum state from a given arbitrary (mixed, in general) state, by simply repeating the same measurement on an ancilla system interacting with the target system \cite{ref:qpf,ref:qpfe}.
Such a mechanism is interesting in itself and is even indispensable when
direct operations on the target quantum systems are not allowed or
unavailable.
The repeated measurements on the ancilla can be regarded as an indirect (POVM) measurement on the target system which, under proper conditions, allows us to  drive the latter toward the desired pure state.
This idea was applied to the initialization of qubits \cite{ref:qpfe}, extraction of entanglement \cite{ref:qpfe,ref:qpfeLidarPaternostroKim} 
and a nonclassical state \cite{ref:qpfDanieleAngular}, and to establish entanglement between separated qubits \cite{ref:qpfeSeparated,ref:qpfescCiccarelloPRL,ref:qpfer,ref:mmm}.

In those schemes, a pure quantum state is obtained from an arbitrary initial configuration, only when the ancilla system is repeatedly confirmed to be in a specific state by all the measurements performed during the protocol.
That is, they are probabilistic schemes characterized by a success probability strictly less than $1$.
The primary motivation of the present work is to persue 
a scheme which would allow one to reach the target state with probability 1, or at least, with probability arbitrarily close to 1, independently of the initial conditions.
To achieve such a goal, we take inspiration from an approach recently introduced in Ref.\ \cite{ref:DanielVittorio}, 
in which a target system is indirectly controlled by making it interact with a sequence of (properly prepared) ancillas, which are then discarded.  
As in the case of Ref.\ \cite{ref:DanielVittorio}, 
our finding relies on a useful property of quantum channels. 
Namely, we make use of the fact that under proper conditions (see, for instance, Refs.\ \cite{ref:GeneLyapunov,ref:Merkli2006} and references therein) repetitive applications of the same map  drive the system toward a fixed point, independently of its initial configuration (\textit{mixing} property).

\begin{figure}[b]
\begin{center}
\includegraphics[width=0.55\textwidth]{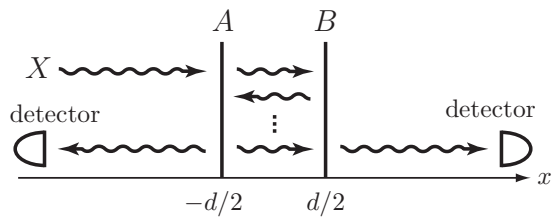}
\caption{A schematic drawing of the setup. An ancilla qubit $X$ is sent to two fixed qubits $A$ and $B$, with a certain wave vector $k$. After being scattered by the delta-shaped potentials produced by $A$ and $B$, we check whether it is reflected or transmitted. 
Neither the preparation of the spin of the incident $X$ nor the spin-resolved detection of the scattered $X$ is required.}
\label{fig:setup}
\end{center}
\end{figure}
As a nontrivial example of such scheme, we concentrate on a prototypical setup which has been extensively investigated in the literature recently \cite{ref:qpfescCiccarelloPRL,ref:qpfer,ref:mmm,ref:qpfeScattering,ref:EntReso,ref:qpfesc-long}. 
Here, two non-interacting  target qubits $A$ and $B$ sit at fixed distance from each other along a 1D channel as sketched in Fig.\ \ref{fig:setup}. 
The goal of the scheme is to drive $A$ and $B$ into an entangled state with the help of a (flying) qubit, which is sent through the 1D channel and is detected after it has been scattered  by the targets.
In the simplest configuration considered so far, the latters are supposed to be initially in a (known) separable state, while the ancilla qubit is prepared in an appropriate spin state before being injected into the setup. 
Under these assumptions, it has been shown that entanglement between $A$ and $B$ can be generated in a probabilistic fashion by a simple post-selection of the spin state of the scattered ancilla \cite{ref:qpfeScattering,ref:EntReso,ref:qpfesc-long}.  
Schemes which do not require preparation of the initial state of the target qubits to a specific state were also proposed, in which entanglement is extracted after repetition of scattering + post-selection \cite{ref:qpfescCiccarelloPRL,ref:qpfer,ref:mmm}.
Such improved protocols however are still probabilistic, as they produce the desired entanglement only with a certain success probability. 
In contrast, the approach we present here 
generates a maximally entangled state between $A$ and  $B$  \textit{from an arbitrary initial state with probability 1}. 
Furthermore, it requires neither the preparation nor the post-selection of the spin state of the ancilla qubits.

The paper is organized as follows.
In Sec.\ \ref{sec:Setup}, we introduce the setup and give some preliminary definitions. 
Section \ref{sec:Protocol} introduces the protocol and states the main result of our work. 
The proof of the latter is then provided in Sec.\ \ref{sec:Proof}. 
Section \ref{sec:Robustness} is devoted to analyze the robustness of the scheme, while conclusions and remarks are given in Sec.\ \ref{sec:Conclusions}.

\section{Setup}
\label{sec:Setup}
Our setup is sketched in Fig.\ \ref{fig:setup}.
Two qubits $A$ and $B$ are fixed at $x=-d/2$ and $d/2$, respectively, along a 1D channel.
They do not interact directly with each other, while we wish to establish entanglement between them.
To do so, 
we send flying ancilla qubits $X$ as ``mediators'' and let them scatter with $A$ and $B$\@.
As in Refs.\ \cite{ref:qpfescCiccarelloPRL,ref:qpfeScattering,ref:qpfer,ref:mmm,ref:EntReso}, we assume the system to be described by the following Hamiltonian:
\begin{eqnarray}
H=\frac{p^2}{2m}
&{}+g(\bm{\sigma}^{(X)}\cdot\bm{\sigma}^{(A)})\delta(x+d/2)
\nonumber\\
&{}+g(\bm{\sigma}^{(X)}\cdot\bm{\sigma}^{(B)})\delta(x-d/2),
\label{eqn:Hamiltonian}
\end{eqnarray}
where $x$ and $p$ are the position and the momentum of $X$ in 1D, the operators $\bm{\sigma}^{(J)}$ ($J=X,A,B$) represent the Pauli operators of the spins, and the potentials produced by $A$ and $B$ are represented by the delta-shaped potentials.
According to Eq.\ (\ref{eqn:Hamiltonian}), the spin of $X$ interacts separately with $A$ and $B$ through the Heisenberg-type coupling during the scattering.
This Hamiltonian has been proposed to effectively model the coupling between   electrons occupying the lowest sub-band and magnetic impurities placed along 
a quasi one-dimensional (1D) wire, such as a semiconductor quantum wire \cite{ref:DattaTransMeso} or a single-wall carbon nanotube \cite{ref:Tans-Nature1997}, where electrons flow.

The particle $X$ is sent from the left with a fixed incident wave vector $k>0$ and scattered by $A$ and $B$\@.
The matrix elements of the scattering operator $S$ are given by \cite{ref:qpfer,ref:qpfesc-long}
\begin{equation}
\bra{k'\zeta'}S\ket{k\zeta}
=e^{-ikd}[\delta(k'-k)\bra{\zeta'}T_k\ket{\zeta}
+\delta(k'+k)\bra{\zeta'}R_k\ket{\zeta}],
\label{eqn:SmatrixElements}
\end{equation}
where $\ket{k}$ is the eigenstate of the momentum operator $p$ of $X$ belonging to its eigenvalue $\hbar k$, and $\ket{\zeta}$ represents a spin state of $XAB$.
The operators $T_k$ and $R_k$ describe the changes provoked in the spin state of $XAB$ when $X$ is transmitted to the right and reflected to the left, respectively.
They satisfy the unitarity condition
\begin{equation}
T_k^\dag T_k+R_k^\dag R_k=\openone_{XAB},
\label{eqn:Unitarity}
\end{equation}
and are given by
\numparts
\begin{eqnarray}
\fl
T_k
=e^{ikd}\,\Bigl[
\alpha_k(1-4 i\Omega_k)P_-
+(\alpha_kQ_\frac{1}{2}+\beta_kQ_\frac{3}{2})P_+
\nonumber\\
\fl\qquad\qquad\qquad\qquad\quad\ \ %
{}-\alpha_k\Omega_k^2(1-e^{2 ikd})
(
P_-
-3Q_\frac{1}{2}P_+
-K_++K_-
)
\Bigr],
\label{eqn:T}
\end{eqnarray}
\begin{eqnarray}
\fl
R_k
=T_k e^{- ikd}-1
-i\Omega_k(1-e^{2 ikd}) 
\,\biggl\{
6\alpha_k \Omega_k^2(1-e^{2 ikd})P_-
+(2\alpha_k Q_\frac{1}{2}
-\beta_k Q_\frac{3}{2})P_+
\nonumber\\
\fl\qquad\qquad\qquad\qquad\quad\ \ %
{}+\frac{1}{2}\alpha_k 
[
1+3\Omega_k^2(1-e^{2 ikd})
-4 i\Omega_k P_+
]
(K_++K_-)
\biggr\},
\label{eqn:R}
\end{eqnarray}
\endnumparts
where
\numparts
\begin{eqnarray}
\fl
\alpha_k =\frac{1}{\displaystyle
(1-4  i\Omega_k) +2 \Omega_k^2(1-6 i\Omega_k) (1-e^{2 ikd})+9\Omega_k^4 (1-e^{2 ikd})^2 
},\\
\fl
\beta_k =\frac{1}{(1+2  i\Omega_k)-\Omega_k^2(1-e^{2 ikd})},\qquad
\Omega_k=\frac{mg}{\hbar^2k}.
\end{eqnarray}
\endnumparts
In the above expressions,
\begin{equation}
P_-
=\frac{1-\bm{\sigma}^{(A)}\cdot\bm{\sigma}^{(B)}}{4}
,\qquad
P_+
=\frac{3+\bm{\sigma}^{(A)}\cdot\bm{\sigma}^{(B)}}{4}
\label{eqn:Ppm}
\end{equation}
are the projection operators on the singlet and triplet sectors of $A$ and $B$, respectively, while
\numparts
\label{eqn:Q}
\begin{eqnarray}
Q_\frac{3}{2}=\frac{2}{3}P_++\frac{1}{6}\bm{\sigma}^{(X)}\cdot(\bm{\sigma}^{(A)}+\bm{\sigma}^{(B)}),
\\
Q_\frac{1}{2}=P_-+\frac{1}{3}P_+-\frac{1}{6}\bm{\sigma}^{(X)}\cdot(\bm{\sigma}^{(A)}+\bm{\sigma}^{(B)})
\end{eqnarray}
\endnumparts
are those on the spin-$\frac{3}{2}$ and spin-$\frac{1}{2}$ sectors of $XAB$, respectively \cite{ref:qpfesc-long} 
(note that they are all commuting with each other and satisfy $Q_\frac{3}{2}P_-=P_-Q_\frac{3}{2}=0$).\footnote{Throughout this paper, the unit operators are often omitted as $\openone_X\otimes P_\pm\to P_\pm$, $\openone_X\otimes\bm{\sigma}^{(A)}\otimes\openone_B\to\bm{\sigma}^{(A)}$, $3\openone_{AB}\to3$, etc.}
The other operators
\begin{equation}
K_\pm=\bm{\sigma}^{(X)}\cdot\bm{\Sigma}_\pm^{(AB)},
\end{equation}
defined with 
\begin{eqnarray}
\bm{\Sigma}_+^{(AB)}
&=\frac{1}{2}[
(\bm{\sigma}^{(A)}-\bm{\sigma}^{(B)})
+ i(\bm{\sigma}^{(A)}\times\bm{\sigma}^{(B)})
]\nonumber
\\
&=P_+(\bm{\sigma}^{(A)}-\bm{\sigma}^{(B)})
=P_+i(\bm{\sigma}^{(A)}\times\bm{\sigma}^{(B)})
\nonumber
\\
&
=(\bm{\sigma}^{(A)}-\bm{\sigma}^{(B)})P_-
=i(\bm{\sigma}^{(A)}\times\bm{\sigma}^{(B)})P_-
\nonumber
\\
&=\bm{\Sigma}_-^{(AB)\dag},
\end{eqnarray}
are responsible for the transitions between the singlet and triplet sectors of $A$ and $B$, with the only nonzero elements $P_\pm\bm{\Sigma}_\pm^{(AB)} P_\mp\neq0$.

\begin{figure}[t]
\begin{center}
\includegraphics{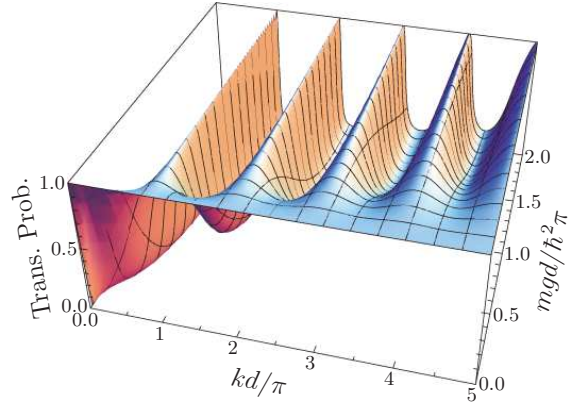}
\caption{Transmission probability of the ancilla qubit $X$ when it is injected from the left with its spin prepared in $\ket{\uparrow}_X$. 
The plot shows the probability of detecting the ancilla on the right detector after the scattering by $A$ and $B$.
Here $A$ and $B$ are assumed to be initially in the singlet state $\ket{\Psi^-}_{AB}$, so that the quantity plotted is nothing but $\Tr\{T_k(\ketbras{\uparrow}{\uparrow}{X}\otimes\ketbras{\Psi^-}{\Psi^-}{AB})T_k^\dag\}$. 
Its dependence on the incident wave vector $k$ of $X$ and the coupling constant $g$ is shown.
It exhibits resonances 
at $k=n\pi/d$ ($n=1,2,\ldots$) for any $g\neq0$.
}
\label{fig:Reso}
\end{center}
\end{figure}
It is pointed out in Ref.\ \cite{ref:EntReso} that this system exhibits interesting resonant transmissions controlled by the entanglement in $A$ and $B$: for instance, when $A$ and $B$ are in the singlet state $\ket{\Psi^-}_{AB}=(\ket{\uparrow\downarrow}_{AB}-\ket{\downarrow\uparrow}_{AB})/\sqrt{2}$, the potentials produced by $A$ and $B$ look ``transparent'' for $X$ sent with a resonant wave vector $k_n$ satisfying the resonance condition $k_nd=n\pi$ ($n=1,2,\ldots$).
See Fig.\ \ref{fig:Reso}.
More explicitly, the transmission and reflection operators $T_k$ and $R_k$ at the resonance points are given by
\numparts
\label{eqn:TRreso}
\begin{eqnarray}
T_{k_n}
=(-1)^n\left[
P_-
+\left(
\frac{Q_\frac{1}{2}}{1-4  i\Omega_{k_n}}
+\frac{Q_\frac{3}{2}}{1+2  i\Omega_{k_n}}
\right)
P_+
\right],
\label{eqn:Treso}
\\
R_{k_n}
=\left(
\frac{4 i\Omega_{k_n}}{1-4  i\Omega_{k_n}}
Q_\frac{1}{2}
-\frac{2 i\Omega_{k_n}}{1+2  i\Omega_{k_n}}
Q_\frac{3}{2}
\right)
P_+,
\label{eqn:Rreso}
\end{eqnarray}
\endnumparts
showing that $X$ is perfectly transmitted without spin flip when $A$ and $B$ are in the singlet state $\ket{\Psi^-}_{AB}$.

\section{Protocol}
\label{sec:Protocol}
To construct our 
scheme, 
we make use of the resonance condition detailed in the previous section. 
Similar approaches have been explored in Refs.\ \cite{ref:qpfescCiccarelloPRL,ref:qpfer,ref:mmm}.
Specifically, in Ref.\ \cite{ref:qpfescCiccarelloPRL}, Ciccarello \textit{et~al.} exploited repetition of the scattering of $X$ with a resonant momentum followed by an appropriate post-selection on $X$, to extract the singlet state $\ket{\Psi^-}_{AB}$ from $AB$ with a probability which depends upon the initial state of the system.
In contrast, in Ref.\ \cite{ref:qpfer,ref:mmm}, a scheme is proposed to extract the singlet state from an arbitrarily given initial state of $AB$, in which neither the preparation of the spin state of $X$ nor its post-selection is required.
However, it is still a probabilistic scheme, since the singlet state is extracted by the successive post-selections of transmitted events.
The protocol presented here solves this problem, allowing one to produce a maximally entangled state of $A$ and $B$ from their arbitrary initial state \textit{with probability 1}.
The scheme remains free from the preparation and the post-selection of the spin state of $X$.

\begin{figure}[t]
\begin{center}
\includegraphics{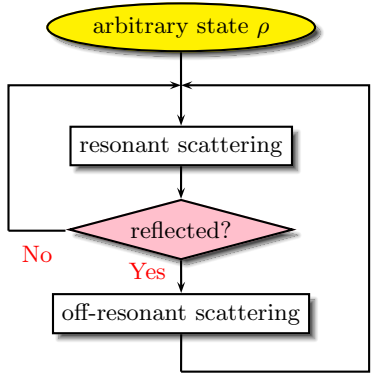}
\caption{Flowchart of the protocol.}
\label{fig:Flowchart}
\end{center}
\end{figure}

Here comes our protocol (see Fig.\ \ref{fig:Flowchart}):
\begin{enumerate}
\renewcommand{\labelenumi}{\arabic{enumi}.}
\setcounter{enumi}{-1}
\item The initial state of $A$ and $B$ is \textit{arbitrary} and is in general a mixed state $\rho$.
\item We send $X$ with its spin \textit{arbitrary} from the left to $A$ and $B$ with a resonant wave vector $k_n=n\pi/d$ ($n=1,2,\ldots$), and see if it is transmitted to the right or reflected to the left, \textit{irrespectively of its spin state}.
\item If $X$ is detected on the right (transmitted), we proceed to the next round (to step 1).
\item Otherwise (reflected), we send another $X$ from the left with its spin \textit{randomly} chosen with an off-resonant wave vector $q$ (perfectly polarized incident spin of $X$ is not recommended).
We do not check anything after this scattering; we just proceed to the next round (to step 1).
\item We repeat this routine (steps 1 to 3) many times and end up with the singlet state $\ket{\Psi^-}_{AB}$ in $A$ and $B$ \textit{with probability 1}.
\end{enumerate}

This is a feedback approach, where we take different actions depending on the outcome of the resonant scattering at step 1. 
Notice however that, for implementing the feedback, no additional element is required: we have only to set the incident wave vector of the ancilla off-resonant.

It is also worth stressing that, at step 1, the choice of the incident spin of $X$ is irrelevant and we can just choose it arbitrarily.
At step 3, on the other hand, we can choose it randomly, but if the incident spin of $X$ is perfectly polarized always in the same direction at every cycle, the scheme does not work.
In order to make the following analysis simpler and transparent, however, we take unpolarized spin for both steps, represented by the completely mixed state $\openone_X/2$.

A single cycle (steps 1 to 3) changes the spin state $\rho$ of $A$ and $B$ in the following way:
\begin{equation}
\rho
\longrightarrow
\mathcal{M}\rho
=(\mathcal{T}_{k_n}+\mathcal{S}_q\mathcal{R}_{k_n})\rho,
\label{eqn:M}
\end{equation}
where
\numparts
\begin{eqnarray}
\mathcal{T}_k\rho
=\Tr_{X}\{T_k(\openone_{X}/2\otimes\rho)T_k^\dag\},\\
\mathcal{R}_k\rho
=\Tr_{X}\{R_k(\openone_{X}/2\otimes\rho)R_k^\dag\},
\end{eqnarray}
\endnumparts
and
\begin{equation}
\mathcal{S}_k=\mathcal{T}_k+\mathcal{R}_k.
\label{eqn:Sk}
\end{equation}
On the basis of the unitarity (\ref{eqn:Unitarity}), the map $\mathcal{M}$ is trace preserving,
\begin{equation}
\Tr_{AB}\{\mathcal{M}\rho\}=1,
\end{equation}
meaning that all possible outcomes are properly kept at each step and there is no selection of specific detection events.

We will see that the repetition of the above cycle leads $A$ and $B$ into the singlet state,
\begin{equation}
\mathcal{M}^N\rho
\to
\ketbras{\Psi^-}{\Psi^-}{AB}
\quad(N\to\infty)
,
\label{eqn:Mixing2Psi}
\end{equation}
irrespectively of their initial state $\rho$, where $N$ is the number of cycles of the protocol. 
In other words, we are going to prove that $\mathcal{M}$ is ``mixing'' \cite{ref:GeneLyapunov} with its fixed point given by $\ketbras{\Psi^-}{\Psi^-}{AB}$.

\section{Proof}
\label{sec:Proof}
In order to verify the claim (\ref{eqn:Mixing2Psi}), we invoke an important result on mixing channels, which states that a CPT (completely positive and trace-preserving) map $\mathcal{M}$ is mixing if it has a unique fixed point that is a pure state, e.g., see Ref.\ \cite{ref:GeneLyapunov}. 
We remind that the fixed points of $\mathcal{M}$ are defined as those input states $\rho_*$ which are left invariant by the action of the channel, i.e., 
\begin{equation} 
\mathcal{M}\rho_*= \rho_*.
\end{equation}
In our case, it is immediate to check that $\ketbras{\Psi^-}{\Psi^-}{AB}$ is a fixed point of $\mathcal{M}$: indeed, as is clear from Eqs.\ (\ref{eqn:Treso}) and (\ref{eqn:Rreso}), $\ketbras{\Psi^-}{\Psi^-}{AB}$ is preserved by $\mathcal{T}_{k_n}$ while $\mathcal{R}_{k_n}$ yields nothing (reflection does not occur with $\ketbras{\Psi^-}{\Psi^-}{AB}$).
Thus, since $\ketbras{\Psi^-}{\Psi^-}{AB}$ is a pure state, it follows that the only thing we need to verify in order to prove the mixing (\ref{eqn:Mixing2Psi}) is that there exists no other fixed point of the channel $\mathcal{M}$.

Assume then that there exists another fixed point $\rho_*$, that is different from $\ketbras{\Psi^-}{\Psi^-}{AB}$.
By definition, it must satisfy the identity
\begin{equation}
\mathcal{M}\rho_*
=(\mathcal{T}_{k_n}+\mathcal{S}_q\mathcal{R}_{k_n})\rho_*
=\rho_*.
\label{eqn:FixedPoint}
\end{equation}
This expression can be simplified by noticing that the following identity holds at resonances:
\begin{equation}
\mathcal{P}_-\mathcal{T}_{k_n}
=\mathcal{P}_-,
\label{eqn:ResoProjection}
\end{equation}
where $\mathcal{P}_\pm$ are the superoperators associated with the projections onto the triplet and singlet subspaces, respectively,  
\begin{equation}
\mathcal{P}_\pm\rho=P_\pm\rho P_\pm.
\label{eqn:SuperProjections}
\end{equation}
Indeed, taking Eq.\ (\ref{eqn:ResoProjection}) into account, a necessary condition for Eq.\ (\ref{eqn:FixedPoint}) is given by
\begin{equation}
\mathcal{P}_-\mathcal{S}_q\mathcal{R}_{k_n}\rho_*=0.
\label{eqn:FixedReduced}
\end{equation}
Look at Eq.\ (\ref{eqn:Rreso}) again: it shows that $\mathcal{R}_{k_n}$ cuts out the singlet sector and acts on the triplet components.
Since we are assuming that $\rho_*$ is different from the singlet state $\ketbras{\Psi^-}{\Psi^-}{AB}$, we surely find some component $\mathcal{R}_{k_n}\rho_*\neq0$ in the triplet sector.
Therefore, if $\mathcal{S}_q$ is such a map that certainly couples any triplet components to the singlet sector, the condition (\ref{eqn:FixedReduced}) is never satisfied except for the singlet state $\ketbras{\Psi^-}{\Psi^-}{AB}$. 
Consequently, by contradiction, the singlet state is proved to be the unique fixed point of the map $\mathcal{M}$.

\begin{figure}[t]
\begin{center}
\includegraphics{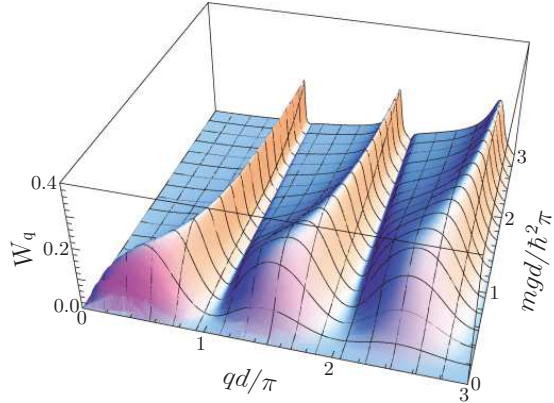}
\caption{The coefficient $W_q$ defined in Eq.\ (\ref{eqn:Wq}), as a function of $q$ and $g$.
The larger is $W_q$, the larger is the flow from the triplet sector to the singlet, by the action of $\mathcal{S}_q$.}
\label{fig:Filter}
\end{center}
\end{figure}
The condition for such a map $\mathcal{S}_q$ is expressed as 
\begin{equation}
\Tr_{AB}\{\mathcal{P}_-\mathcal{S}_q\mathcal{P}_+\rho\}
>0,
\end{equation}
for any state $\rho\neq\ketbras{\Psi^-}{\Psi^-}{AB}$. 
By inserting the explicit expressions of $T_q$ and $R_q$ given in Eqs.\ (\ref{eqn:T}) and (\ref{eqn:R}), it reads
\begin{equation}
\Tr_{AB}\{\mathcal{P}_-\mathcal{S}_q\mathcal{P}_+\rho\}
=W_q\Tr_{AB}\{P_+\rho\},
\end{equation}
with a coefficient
\begin{eqnarray}
W_q
={}&|\alpha_q|^2\Omega_q^2|1-e^{2iqd}|^2
\,\Bigl[
4\Omega_q^2
+|1-2i\Omega_q+3\Omega_q^2(1-e^{2iqd})|^2
\Bigr],
\label{eqn:Wq}
\end{eqnarray}
which is nonvanishing for any $q\neq n\pi/d$ ($n=0,1,2,\ldots$) and for any state $\rho\neq\ketbras{\Psi^-}{\Psi^-}{AB}$.
That is, $\mathcal{S}_q$ with an off-resonant wave vector $q$ is a map that surely couples any triplet component of $A$ and $B$ to the singlet sector and excludes the existence of $\rho_*$ satisfying Eq.\ (\ref{eqn:FixedReduced}) and different from the singlet state $\ketbras{\Psi^-}{\Psi^-}{AB}$.
See Fig.\ \ref{fig:Filter}, where the coefficient $W_q$ is plotted as a function of $q$ and $g$.
They are actually nonzero except at the resonant wave vectors  $q=n\pi/d$ ($n=1,2,\ldots$).

The reason why $A$ and $B$ are attracted into the singlet state by the repeated applications of $\mathcal{M}$ is the following.
The singlet state $\ketbras{\Psi^-}{\Psi^-}{AB}$ is the fixed point of the map $\mathcal{M}$, and the singlet component in the state of $A$ and $B$ remains there.
As for the triplet components, they provoke the reflection of $X$ with a certain probability at step 1.
Once $X$ is found to be reflected, the qubits $A$ and $B$ are ``shuffled'' by the subsequent off-resonant scattering $\mathcal{S}_q$, which creates a singlet component.
In total, the probability to find $A$ and $B$ in the singlet state is increased by the single cycle.
Outflow of the probability from the singlet sector is absent, while inflow is present.
This feature leads the system into the singlet state $\ketbras{\Psi^-}{\Psi^-}{AB}$.\footnote{One may wonder that a reflection event at step 1 suddenly projects out the singlet component and keeps $A$ and $B$ from approaching the singlet state.
As the cycle is repeated, however, the average probability to find $A$ and $B$ in the singlet state increases (as explained above), and the average probability for such a reflection to occur is accordingly reduced: the chance of the loss of the singlet component becomes less and less likely, and the singlet component keeps on growing on average. A simple statistical argument can then be used to claim that, since on average these events are suppressed, the probability of generating a specific trajectory which does not have this property is also suppressed.}

\begin{figure}[t]
\begin{center}
\includegraphics[width=0.55\textwidth]{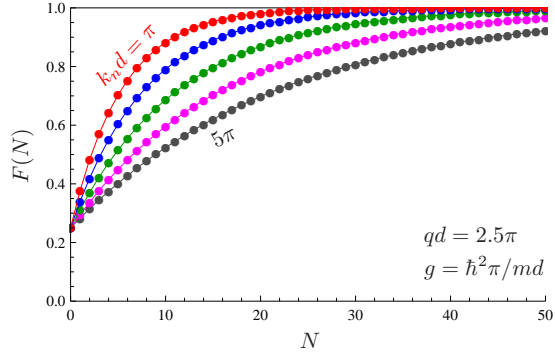}
\caption{Average fidelity $F(N)$ defined in Eq.\ (\ref{eqn:Fidelity}) as a function of the number $N$ of protocol cycles. The five curves refer to different choices of the resonant wave vectors, namely $k_nd/\pi =1,\,2,\,3,\,4,\,5$ from top to bottom.
The other parameters are $qd/\pi=2.5$ and $g=\hbar^2\pi/md$. 
}
\label{fig:Fmixing}
\end{center}
\end{figure}

\begin{figure}
\begin{tabular}{l@{\quad\ \ }l}
\footnotesize(a)&\footnotesize(b)\\[-3.5truemm]
\includegraphics[width=0.46\textwidth]{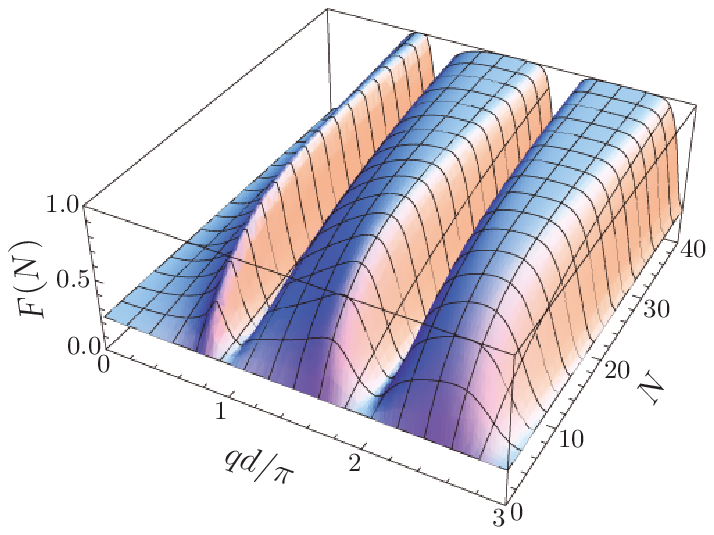}&
\includegraphics[width=0.46\textwidth]{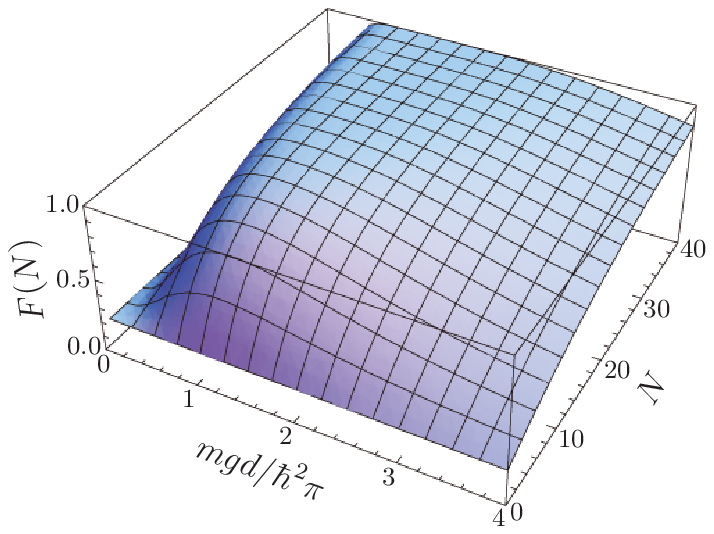}
\end{tabular}
\caption{Average fidelity $F(N)$ defined in Eq.\ (\ref{eqn:Fidelity}). 
(a) Dependence on $q$ with $k_n=\pi/d$ and $g=\hbar^2\pi/md$.
(b) Dependence on $g$ with $k_n=\pi/d$ and $q=2.5\pi/d$.
}
\label{fig:FmixingQG}
\end{figure}

The scheme surely works.
As a direct check, the average fidelity $F(N)$ of the protocol is shown in Figs.\ \ref{fig:Fmixing} and \ref{fig:FmixingQG}. 
It is obtained by computing the fidelity between the generated state $\mathcal{M}^N\rho$ and the target $\ketbras{\Psi^-}{\Psi^-}{AB}$, which is averaged over all possible choices of the input state $\rho$, i.e., 
\begin{equation}
F(N) \equiv \overline{\Tr_{AB}\{ \mathcal{P}_-\mathcal{M}^N \rho\}}
= \Tr_{AB}\{ \mathcal{P}_- \mathcal{M}^N(\openone_{AB}/4)\},
\label{eqn:Fidelity}
\end{equation}
where $\overline{{}\cdots{}\vphantom{]}}$ 
stands for the average over the input state $\rho$ and $\openone_{AB}$ is the identity operator on $AB$. 
From these plots, it is evident that as the number $N$ of protocol cycles increases, the average fidelity $F(N)$ approaches asymptotically to $1$, as long as the wave vector $q$ is properly set off-resonant.
This implies that for an average choice of the input state $\rho$, the state $\mathcal{M}^N\rho$ approaches $\ketbras{\Psi^-}{\Psi^-}{AB}$. 
By linearity, this also implies that the same result should hold for each input $\rho$.\footnote{As a matter of fact, one can easily verify that in case of pure fixed point, asymptotically optimal average fidelity is a sufficient condition for the mixing property of a CPT map.} 
Therefore, Figs.\ \ref{fig:Fmixing} and \ref{fig:FmixingQG} provide an alternative proof of the  mixing property of $\mathcal{M}$.

The speed of the convergence is controlled by the following two factors:
(i) the reflection of $X$ sent with a resonant wave vector $k_n=n\pi/d$ ($n=1,2,\ldots$), at step 1, due to the presence of triplet components in $AB$, and (ii) the transition of $AB$ from the triplet sector to the singlet by the off-resonant scattering with $q\neq n\pi/d$ ($n=1,2,\ldots$), at step 3.
The resonant scattering does not bring any triplet components of $AB$ to the singlet sector at step 1. 
However, the reflection of the resonant $X$ triggers us to proceed to the off-resonant scattering at step 3, which brings some triplet components of $AB$ to the singlet sector.
Therefore, the larger is the reflection probability of the resonant $X$ and the stronger is the transition from the triplet sector to the singlet by the off-resonant scattering, the faster is the convergence to the fixed point.
The former is controlled by $R_{k_n}$ in Eq.\ (\ref{eqn:Rreso}) and the latter by $W_q$ in Eq.\ (\ref{eqn:Wq}). 
In particular, the reflection probability is smaller for a higher incident wave vector $k_n$. 
See $\Omega_{k_n}$ in the numerator of $R_{k_n}$ in Eq.\ (\ref{eqn:Rreso}). 
$W_q$ in Eq.\ (\ref{eqn:Wq}) is also proportional to $\Omega_q^2$ and is a decreasing function of $q$ (for not too small $q$) (apart from the oscillation due to resonance).
The convergence is thus slower with higher incident momenta, as demonstrated in Fig.\ \ref{fig:Fmixing}.

\section{Robustness}
\label{sec:Robustness}
In this section, we analyze the robustness of the proposed scheme. 
We start by considering errors in the preparation of the ancilla qubits. 
Later, we analyze the effect of inefficient detectors.

\subsection{Errors in the incident momentum}
A key ingredient of our entanglement protocol is the resonant tunneling condition of the flying ancillas we impose at step 1.
To see what happens if one fails to enforce such constraint, we consider two alternative scenarios. 
First, we analyze the case in which the source producing the ancilla qubits $X$ is affected by 
  a systematic error that forces it to produce a monochromatic sequence of particles which enters the 1D channel 
 with a constant wave vector $k$, which is not resonant (error by deviation from resonance). 
Then, we consider the situation in which the same source is affected by fluctuations which prevent it from producing monochromatic signals (error by fluctuation of the incident momenta).  
For both scenarios, we compute the overlap between the final state of $AB$ after $N$ protocol cycles, and the target singlet state. 
As one might expect, the case of the systematic error is much more detrimental for the performances of the protocol, with average fidelities which drops below  $50\%$ 
already for small deviations of the impinging momenta. 
On the other hand, the scheme appears to be more resilient to fluctuation errors.

\subsubsection{Deviation from a resonance point}
\begin{figure}[t]
\begin{center}
\includegraphics[width=0.55\textwidth]{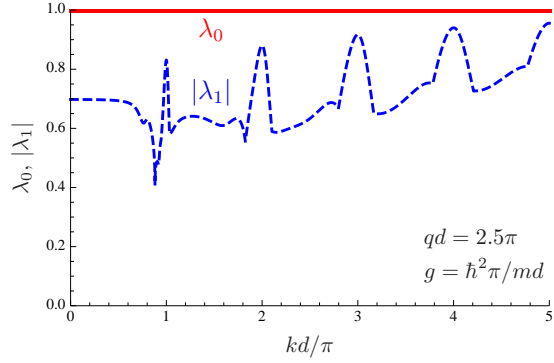}
\caption{The magnitudes of the largest and the second largest eigenvalues, $\lambda_0$ and $\lambda_1$, of the map $\mathcal{M}_{k,q}$ as a function of $k$ with $q=2.5\pi/d$ and $g=\hbar^2\pi/md$.
The presence of a gap between them is a necessary and sufficient condition for mixing \cite{ref:GeneLyapunov}.}
\label{fig:EigenvaluesM}
\end{center}
\end{figure}

Assume that, at step 1 of the protocol, the ancilla qubits $X$ enter the 1D channel with fixed wave vector $k$, which is not necessarily at resonance. Following the derivation in Sec.\ \ref{sec:Protocol}, one can easily verify that, after each protocol cycle, the transformation of the state of $AB$ can still be described as in Eq.\ (\ref{eqn:M}), but with the superoperator $\mathcal{M}$ replaced by the CPT map
\begin{equation}
\mathcal{M}_{k,q}=\mathcal{T}_k+\mathcal{S}_q\mathcal{R}_k.
\label{eqn:Mkq}
\end{equation}
Ideally, $k$ should be set at a resonance $k_n=n\pi/d$ ($n=1,2,\ldots$), while $q$ should be off-resonant. 
As shown in the previous section, these assumptions 
are sufficient to guarantee that 
 the map $\mathcal{M}=\mathcal{M}_{k_n,q}$ is  mixing
 with the singlet state as its fixed point.  
For $k\neq k_n$, however, this is not necessarily true, posing the 
problem on how to compute the state of $AB$ in the asymptotic limit of large $N$ (if $\mathcal{M}_{k,q}$ is not mixing, $\lim_{N\rightarrow \infty
} \mathcal{M}_{k,q}^N \rho$ might not be well defined with the system continuously oscillating between different configurations \cite{ref:GeneLyapunov}). 
For the sake of simplicity, however, we will neglect this issue in the following, assuming the mixing property to hold anyway.   
Even though we do not have a formal proof of this property, such assumption is strongly supported by a series of 
numerical and theoretical evidences. 
We remind in fact that the mixing property of a CPT map is ultimately related with its spectrum \cite{ref:GeneLyapunov}. 
Specifically, a necessary and sufficient condition for mixing is the    existence of a finite gap between the largest 
and second largest eigenvalues of the map. 
As shown in Fig.\ \ref{fig:EigenvaluesM}, this seems to be the case for $\mathcal{M}_{k,q}$.
 Furthermore, since
$\mathcal{M}_{k,q}$ depends continuously on $k$, its spectrum is a continuous function of $k$, and hence, so is its mixing property.
That is, there is at least a neighborhood of $k_n=n\pi/d$ ($n=1,2,\ldots$) such that all $k\in(k_n-\varepsilon,k_n+\varepsilon)$ give rise to mixing maps.
Finally, it is known that the set of non-mixing channels form a subset of zero-measure in the set of CPT maps, so it is highly unlikely to have  $\mathcal{M}_{k,q}$ non-mixing \cite{ref:DanielVittorio}.

\begin{figure}[b]
\begin{center}
\includegraphics[width=0.55\textwidth]{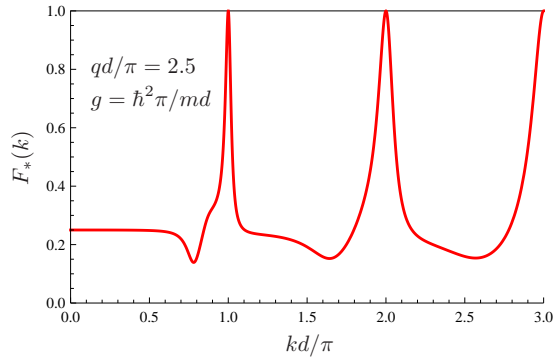}
\caption{Fidelity $F_*(k)$ of the fixed point of the map $\mathcal{M}_{k,q}$ as a function of $k$ for $q=2.5\pi/d$ and $g=\hbar^2\pi/md$.}
\label{fig:F*}
\end{center}
\end{figure}

With the above considerations in mind, we identify the  state of $AB$ after $N\gg1$ protocol cycles for generic $k$ with the fixed point $\rho_*(k)$ of the map $\mathcal{M}_{k,q}$, i.e.,
\begin{equation}
\mathcal{M}_{k,q}\rho_*(k)=\rho_*(k).
\label{eqn:Rho*k}
\end{equation}
Interestingly enough, even without solving Eq.\ (\ref{eqn:Rho*k}) explicitly, it
is possible to derive a concise formula for its fidelity with respect to the singlet state, 
\begin{equation}
F_*(k)\equiv \Tr_{AB}\{\mathcal{P}_-\rho_*(k)\}.
\label{eqn:FidelityK}
\end{equation}
To see this, we first notice
that the transitions between the singlet and triplet sectors of $A$ and $B$ induced by a single scattering event are described by the following $2\times2$ matrices: when $X$ is transmitted,
\begin{equation}
\left(\begin{array}{c}
\medskip
\displaystyle
\Tr_{AB}\{\mathcal{P}_-\mathcal{T}_k\rho\}\\
\displaystyle
\Tr_{AB}\{\mathcal{P}_+\mathcal{T}_k\rho\}
\end{array}\right)
=\left(\begin{array}{cc}
\medskip
\displaystyle
\mathcal{T}_k^{--}&
\mathcal{T}_k^{-+}\\
\mathcal{T}_k^{+-}&
\mathcal{T}_k^{++}
\end{array}\right)
\left(\begin{array}{c}
\medskip
\displaystyle
\Tr_{AB}\{\mathcal{P}_-\rho\}\\
\displaystyle
\Tr_{AB}\{\mathcal{P}_+\rho\}
\end{array}\right)
\end{equation}
with
\numparts
\begin{eqnarray}
\mathcal{T}_k^{--}
={}&|\alpha_k|^2
|
1-4i\Omega_k-\Omega_k^2(1-e^{2ikd})
|^2,
\\
\mathcal{T}_k^{-+}
={}&\frac{1}{3}\mathcal{T}_k^{+-}
=4|\alpha_k|^2\Omega_k^4
|
1-e^{2ikd}
|^2,
\\
\mathcal{T}_k^{++}
={}&\frac{1}{9}|
(\alpha_k+2\beta_k)
+3\alpha_k\Omega_k^2(1-e^{2ikd})
|^2
\nonumber\\
&{}+\frac{2}{9}|
(\alpha_k-\beta_k)
+3\alpha_k\Omega_k^2(1-e^{2ikd})
|^2,
\end{eqnarray}
\endnumparts
and similarly, when $X$ is reflected, with
\numparts
\begin{eqnarray}
\fl
\mathcal{R}_k^{--}
=
|
1-\alpha_k(1-4i\Omega_k)
+\alpha_k\Omega_k^2(1-e^{2ikd})
+6i\alpha_k\Omega_k^3(1-e^{2ikd})^2
|^2,
\\
\fl
\mathcal{R}_k^{-+}
=\frac{1}{3}\mathcal{R}_k^{+-}
=|\alpha_k|^2\Omega_k^2
|
1-e^{2ikd}
|^2
|
1-2i\Omega_k
+3\Omega_k^2(1-e^{2ikd})
|^2,
\\
\fl
\mathcal{R}_k^{++}
=\frac{1}{9}|
3-(\alpha_k+2\beta_k)
+i\Omega_k[2(\alpha_k-\beta_k)+3i\alpha_k\Omega_k](1-e^{2ikd})
|^2
\nonumber\\
\fl\qquad\qquad
{}+\frac{2}{9}|
(\alpha_k-\beta_k)
-i\Omega_k[(2\alpha_k+\beta_k)+3i\alpha_k\Omega_k](1-e^{2ikd})
|^2.
\end{eqnarray}
\endnumparts
In this notation, $W_q$ defined in Eq.\ (\ref{eqn:Wq}) is expressed as $W_q=\mathcal{T}_q^{-+}+\mathcal{R}_q^{-+}$, and the $2\times2$ matrix for $\mathcal{S}_q$ reads
\begin{equation}
\mathcal{S}_q
=\left(\begin{array}{cc}
\medskip
1-3W_q&W_q
\\
3W_q&1-W_q
\label{eqn:Smatrix}
\end{array}\right).
\end{equation}
[By abuse of notation, we use the same symbols for the corresponding $2\times2$ matrices, e.g., $\mathcal{S}_q$ for the $2\times2$ matrix in Eq.\ (\ref{eqn:Smatrix}).]
Combining these expressions, the matrix for $\mathcal{M}_{k,q}$ is also constructed according to Eq.\ (\ref{eqn:Mkq}).
In particular, at resonances $k_n=n\pi/d$ ($n=1,2,\ldots$), we have
\begin{equation}
\mathcal{T}_{k_n}
=\left(\begin{array}{cc}
\medskip
1&0\\
0&1-V_n
\end{array}\right),
\label{eqn:TmatrixReso}
\qquad
\mathcal{R}_{k_n}
=\left(\begin{array}{cc}
\medskip
0&0\\
0&V_n
\end{array}\right)
\end{equation}
with
\begin{equation}
V_n=\frac{8\Omega_{k_n}^2(1+8\Omega_{k_n}^2)}{(1+16\Omega_{k_n}^2)(1+4\Omega_{k_n}^2)},
\label{eqn:Vn}
\end{equation}
and for the ideal map, 
\begin{equation}
\mathcal{M}=\mathcal{M}_{k_n,q}
=\left(\begin{array}{cc}
\medskip
1&W_qV_n
\\
0&1-W_qV_n
\label{eqn:Mmatrix}
\end{array}\right).
\end{equation}
Now, by noting Eq.\ (\ref{eqn:Rho*k}), $\mathcal{P}_+=1-\mathcal{P}_-$, and $\Tr_{AB}\rho_*(k)=1$, the definition of the fidelity (\ref{eqn:FidelityK}) is arranged as
\begin{eqnarray}
F_*(k)
&=\Tr_{AB}\{\mathcal{P}_-\rho_*(k)\}
\nonumber\\
&=\Tr_{AB}\{\mathcal{P}_-\mathcal{M}_{k,q}\rho_*(k)\}
\nonumber
\\
&=\mathcal{M}_{k,q}^{--}\Tr_{AB}\{\mathcal{P}_-\rho_*(k)\}
+\mathcal{M}_{k,q}^{-+}\Tr_{AB}\{\mathcal{P}_+\rho_*(k)\}
\nonumber
\\
&=\mathcal{M}_{k,q}^{-+}
+(\mathcal{M}_{k,q}^{--}-\mathcal{M}_{k,q}^{-+})\Tr_{AB}\{\mathcal{P}_-\rho_*(k)\}
\nonumber
\\
&=\mathcal{M}_{k,q}^{-+}
+(\mathcal{M}_{k,q}^{--}-\mathcal{M}_{k,q}^{-+})F_*(k).
\end{eqnarray}
Therefore, as long as $1-\mathcal{M}_{k,q}^{--}+\mathcal{M}_{k,q}^{-+}>0$ (which is assured by $\mathcal{M}_{k,q}^{-+}>0$ or $\mathcal{M}_{k,q}^{--}<1$), one gets a concise formula for the fidelity,
\begin{equation}
F_*(k)=\frac{\mathcal{M}_{k,q}^{-+}}{1-\mathcal{M}_{k,q}^{--}+\mathcal{M}_{k,q}^{-+}}.
\label{eqn:F*}
\end{equation}
In Fig.\ \ref{fig:F*}, we report its plot as a function of the incident wave vector $k$ for a fixed $q$: as anticipated, 
the fidelity $F_*(k)$ drops below $50\%$ as $k$ deviates from a resonance point $k_n=n\pi/d$ ($n=1,2,\ldots$) by less than a percent.

\subsubsection{Wave packet or fluctuation of the incident momenta}
Consider now the case in which the source emitting $X$ injects a stream of non-monochromatic ancillas into the 1D channel.
Specifically,
let $\psi(k)$ and $\phi(q)$ represent the wave packets in momentum space of the particles produced by the source at steps 1 and  3 of the protocol, respectively.
Then, the map $\mathcal{M}_{k,q}$ of the previous section is substituted by\footnote{We assume that the wave packets are composed only of positive momenta. 
It is possible to show however that the presence of the negative momenta does not spoil the mixing property to be argued in this section. 
In any case, it should be reasonable to assume that such components are negligibly small.} 
\begin{equation}
\tilde{\mathcal{M}}
=\int_0^\infty dk\int_0^\infty dq\,|\psi(k)|^2|\phi(q)|^2\mathcal{M}_{k,q}.
\label{eqn:MP}
\end{equation}
Note that the trace over the momentum degrees of freedom is taken since the detectors do not resolve the momenta, and as a result, only the diagonal components with respect to the momenta contribute to the formula.
This implies that a different type of fluctuation in momentum, incoherent fluctuation, is described by formally the same formula as Eq.\ (\ref{eqn:MP}).
Indeed suppose 
that the incident wave vectors $k$ and $q$ differ from run to run.
Then, the state generated after $N$ cycles of the protocol reads
\begin{equation}
\rho_{(k_1,q_1),\ldots,(k_N,q_N)}
=\mathcal{M}_{k_N,q_N}\cdots\mathcal{M}_{k_1,q_1}\rho.
\label{eqn:SampleRho}
\end{equation}
If the fluctuations of the mementa $(k_i,q_i)$ at each cycle $i=1,\ldots, N$ are characterized by a common probability distribution function $f(k_i,q_i)$, the state (\ref{eqn:SampleRho}) averaged over the probability distribution reads
\begin{equation}
\tilde{\rho}(N)
=\int\biggl(\prod_{i=1}^Ndk_i\,dq_i\,f(k_i,q_i)\biggr)\,
\rho_{(k_1,q_1),\ldots,(k_N,q_N)}
=\tilde{\mathcal{M}}^N\rho,
\end{equation}
where in this case
\begin{equation}
\tilde{\mathcal{M}}
=\int_0^\infty dk\int_0^\infty dq\,f(k,q)\mathcal{M}_{k,q},
\label{eqn:MF}
\end{equation}
which coincides with the expression of Eq.\ (\ref{eqn:MP}) by identifying $f(k,q)$ with $|\psi(k)|^2|\phi(q)|^2$.

It is worth stressing that, differently from the case treated in the previous subsection, one can show that the average map $\tilde{\mathcal{M}}$ is mixing, provided that the distribution $f(k,q)$ overlaps with a resonant wave vector in $k$ and has only measure zero in $q$ at resonances. 
Indeed, split $\tilde{\mathcal{M}}$ into two parts as
\begin{equation}
\tilde{\mathcal{M}}
=\int_{k_n-\varepsilon}^{k_n+\varepsilon}dk
\int_0^\infty dq\,f(k,q)\mathcal{M}_{k,q}
+\tilde{\mathcal{M}}'.
\label{eqn:TildeMpart}
\end{equation}
Recall then that any nontrivial convex sum of a mixing map $\mathcal{E}$ and something else $\mathcal{E}'$ (not necessarily mixing),
\begin{equation}
\tilde{\mathcal{E}}
=\lambda\mathcal{E}
+(1-\lambda)\mathcal{E}'
\quad(0<\lambda\le1),
\label{eqn:ConvexMixing}
\end{equation}
is also a mixing map \cite{ref:DanielVittorio}.
Since the first part of Eq.\ (\ref{eqn:TildeMpart}) is mixing, and has measure nonzero, this theorem ensures that $\tilde{\mathcal{M}}$ is also mixing, implying that the mixing property of $\tilde{\mathcal{M}}$ is robust against the momentum fluctuation. 
As a consequence, in the limit of infinitely many protocol cycles, the system $AB$ is driven to the fixed point $\tilde{\rho}_*$ of the channel $\tilde{\mathcal{M}}$ in Eq.\ (\ref{eqn:MF}). 
Its fidelity with respect to the singlet state can now be computed similarly to Eq.\ (\ref{eqn:F*}), yielding
\begin{equation}
\tilde{F}_*\equiv \Tr_{AB}\{\mathcal{P}_-\tilde{\rho}_*\} = 
\frac{\tilde{\mathcal{M}}^{-+}}{1-\tilde{\mathcal{M}}^{--}+\tilde{\mathcal{M}}^{-+}},
\label{eqn:F*F}
\end{equation}
where the matrix elements $\tilde{\mathcal{M}}^{-\pm}$ are obtained by averaging $\mathcal{M}_{k,q}^{-\pm}$ over the distribution $f(k,q)$.

\begin{figure}
\begin{tabular}{l@{\quad\ \ }l}
\footnotesize(a)&\footnotesize(b)\\[-3.5truemm]
\includegraphics[width=0.46\textwidth]{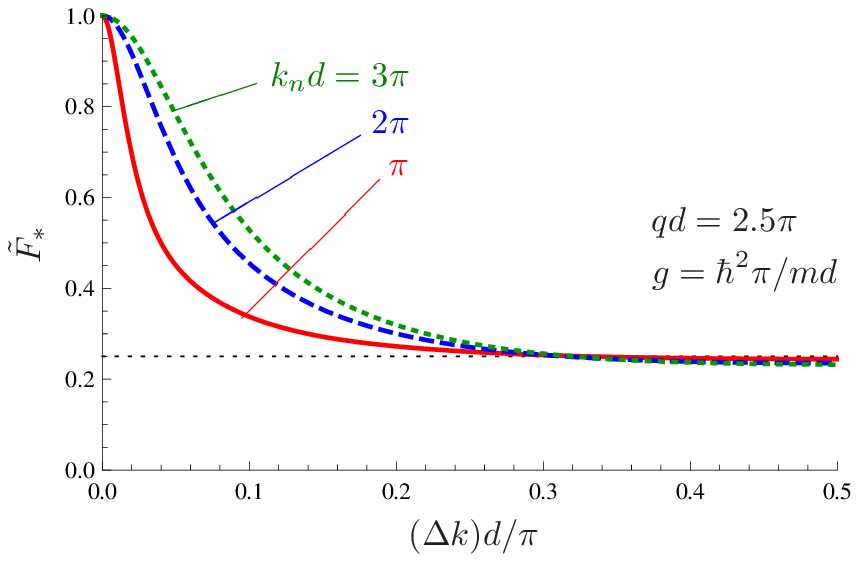}&
\includegraphics[width=0.46\textwidth]{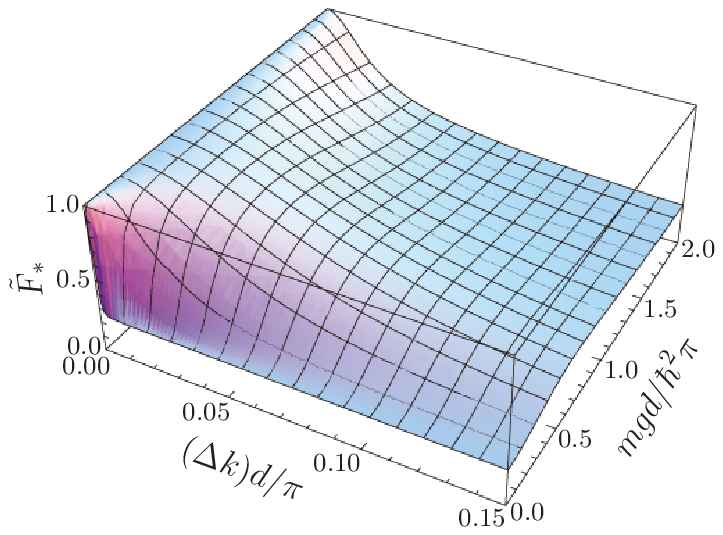}
\end{tabular}
\caption{Average fidelity $\tilde{F}_*$ given in Eq.\ (\ref{eqn:F*F}) of the fixed point of the map $\tilde{\mathcal{M}}$ for a Gaussian distribution of $k$ in Eq.\ (\ref{eqn:Gaussian}), 
(a) as a function of $\Delta k$ for different resonant wave vectors $k_nd/\pi=1$ (solid), $2$ (dashed), $3$ (dotted), with $qd/\pi=2.5$ and $g=\hbar^2\pi/md$, and 
(b) as a function of $\Delta k$ and $g$ with $k_nd/\pi=1$ and $qd/\pi=2.5$.
}
\label{fig:FDKG}
\end{figure}

\begin{figure}
\begin{center}
\includegraphics[width=0.55\textwidth]{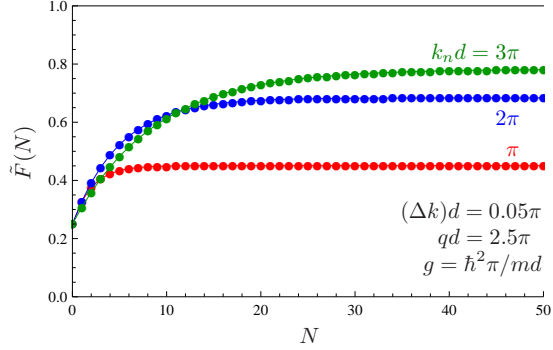}
\caption{Average fidelity $\tilde{F}(N)=\Tr_{AB}\{\mathcal{P}_-\tilde{\mathcal{M}}^N(\openone_{AB}/4)\}$ as a function of the number $N$ of protocol cycles, for a Gaussian distribution of $k$ in Eq.\ (\ref{eqn:Gaussian}). The three curves refer to different choices of the resonant wave vectors, namely $k_nd/\pi =1,\,2,\,3$, with the other parameters fixed at $(\Delta k)d/\pi=0.05$, $qd/\pi=2.5$, and $g=\hbar^2\pi/md$. 
}
\label{fig:FmixingF}
\end{center}
\end{figure}

In Fig.\ \ref{fig:FDKG}, the fidelity $\tilde{F}_*$ is plotted for a Gaussian distribution of $k$, centered at a resonance $k_n$ with width $\Delta k$, for a fixed $q$,
\begin{equation}
f(k,q)\propto\theta(k)e^{-(k-k_n)^2/2(\Delta k)^2}
\delta(q-\bar{q}),
\label{eqn:Gaussian}
\end{equation}
which is normalized as $\int_0^\infty dk\int_0^\infty dq\,f(k,q)=1$.
As expected, the fidelity $\tilde{F}_*$ decreases from unity, as the width of the distribution $\Delta k$ grows.
Notably, in this case, the scheme seems to be quite resilient to the noise: the fidelity drops below $50\%$ only for $\Delta k/k_n$ of the order of $5\%$. 
Furthermore, the decrease is slower than linear for sufficiently small $\Delta k$, and it is less pronounced  
for larger central resonant wave vector $k_n$ and a smaller (but not too small) coupling constant $g$ (although the approach to the final state is slower with a larger $k_n$, as demonstrated in Fig.\ \ref{fig:FmixingF}).

\subsection{Detector efficiency}\label{sec:DE}
In this section, we analyze how defective detections of the scattered ancillas deteriorate the performances of the protocol.
In particular, we consider the case in which the detectors in Fig.\ \ref{fig:setup} are characterized by an efficiency $\eta<1$, i.e.,  
they fail to report the arrival of a particle $X$ with probability $1-\eta$ (for the sake of simplicity, we assume that the two detectors have the same efficiency $\eta$).

To account for such events, we need to specify the action one has to take 
when the detectors fail to report the arrival of a scattered particle
incident on resonance at step 1 of the protocol (since we do not check anything after the scattering of an off-resonant particle at step 3, the efficiency of the detector does not matter for this step).
Specifically, we analyze two alternative solutions.
\begin{figure}[b]
\begin{center}
\begin{tabular}{l@{\qquad}l}
\footnotesize(a)&\footnotesize(b)\\[-3.5truemm]
\includegraphics{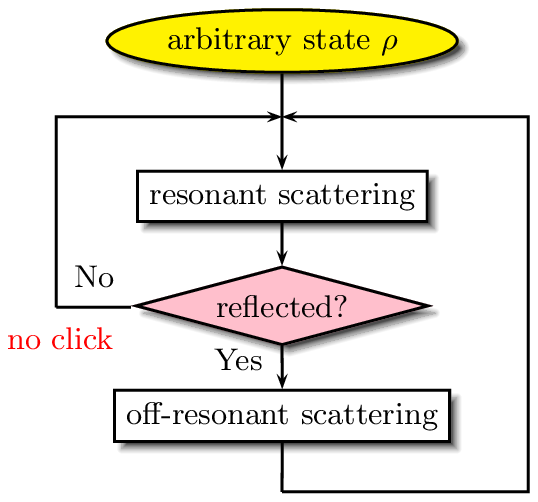}
&\quad\includegraphics{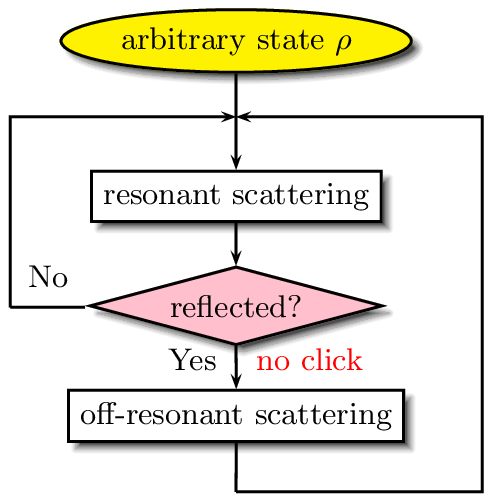}
\end{tabular}
\caption{Flowcharts for Cases I (a) and II (b).}
\label{fig:FlowchartEta}
\end{center}
\end{figure}

\begin{figure}
\begin{center}
\includegraphics{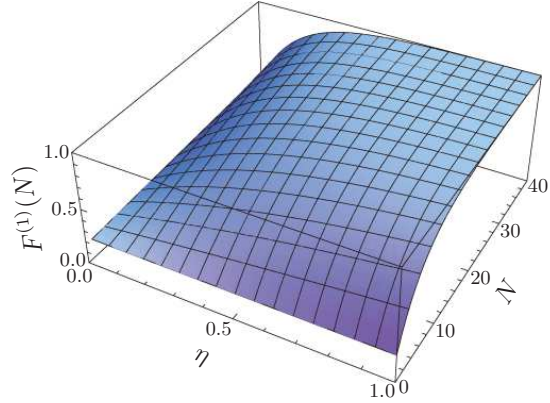}
\caption{Plot of the average fidelity $F^{(1)}(N)=\Tr_{AB}\{\mathcal{P}_-\mathcal{M}_\eta^{(1)N}(\openone_{AB}/4)\}$ 
for inefficient detectors, with the strategy described in Case I of Sec.\ \ref{sec:DE}. 
$N$ is the number of
protocol cycles, while $\eta$ is the detector efficiency. 
The other parameters are $k_n=\pi/d$, $q=2.5\pi/d$, and $g=\hbar^2\pi/md$.
}
\label{fig:newfig}
\end{center}
\end{figure}

\textit{Case I:}\ \ %
We may simply proceed to the next round (step 1) to send the next particle on resonance [Fig.\ \ref{fig:FlowchartEta}(a)].
In such a case, the map $\mathcal{M}$ is modified to 
\begin{equation}
\mathcal{M}_\eta^{(1)}
=\eta\mathcal{M}+(1-\eta)\mathcal{S}_{k_n},
\end{equation}
which is still mixing by the same argument for Eq.\ (\ref{eqn:TildeMpart}).
By noting the expressions in 
Eqs.\ (\ref{eqn:Smatrix}) and (\ref{eqn:Mmatrix}), its $2\times2$ matrix which describes the transitions between the singlet and triplet sectors reads 
\begin{equation}
\mathcal{M}_\eta^{(1)}
=\left(\begin{array}{cc}
\medskip
1&\eta W_qV_n
\\
0&1-\eta W_qV_n
\label{eqn:Mmatrix1}
\end{array}\right),
\end{equation}
and by applying the formula (\ref{eqn:F*}), the fidelity of its fixed point to the target singlet state is shown to remain 
\begin{equation}
F_*^{(1)}(\eta)=1.
\end{equation}
Therefore, the protocol is still able to extract the singlet state from $AB$. 
Notice however that the speed of the convergence of the scheme is affected by $\eta<1$, as shown in Fig.\ \ref{fig:newfig}.
This is a consequence of the fact that, in this case, the element of the scattering matrix (\ref{eqn:Mmatrix1}) associated with the transition from the triplet sector to the singlet gets degraded: the smaller is $\eta$, the slower is the speed of convergence.

\textit{Case II:}\ \ %
We can proceed to step 3 (as if the ancilla $X$ on resonance is reflected) to send $X$ with the off-resonant wave vector $q$ [Fig.\ \ref{fig:FlowchartEta}(b)].
In this case, the map $\mathcal{M}$ is changed to 
\begin{equation}
\mathcal{M}_\eta^{(2)}
=\eta\mathcal{M}+(1-\eta)\mathcal{S}_q\mathcal{S}_{k_n},
\end{equation}
which is also mixing.
It yields
\begin{equation}
\mathcal{M}_\eta^{(2)}
=\left(\begin{array}{cc}
\medskip
1-3(1-\eta)W_q&
(1-\eta+\eta V_n)W_q\\
3(1-\eta)W_q&
1-(1-\eta+\eta V_n)W_q
\end{array}\right),
\end{equation}
and the fidelity of its fixed point is given by
\begin{equation}
F_*^{(2)}(\eta)
=\frac{(1-\eta)+\eta V_n}{4(1-\eta)+\eta V_n},
\end{equation}
which ranges between $0.25$ and $1$. 
This strategy is thus not as effective as the previous one in terms of fidelity.
This is because the off-resonant scattering in the absence of guarantee that there is no singlet component in $AB$, which should be assured by the detection of a reflected particle on resonance, provokes the undesired transition from the singlet sector to the triplet.
The presence of this leakage channel hinders the convergence to the singlet state.

\section{Conclusions and remarks}
\label{sec:Conclusions}
We have proposed and studied a scheme for preparing a maximally entangled state in two non-interacting qubits, initially given in an arbitrary state.
By repetition of resonant scattering of ancilla qubit, followed by off-resonant scattering if necessary, qubits $A$ and $B$ are driven from any initial state into the singlet state with probability 1 (perfect efficiency).
Neither the preparation nor the post-selection of the ancilla spin state is required.
By introducing an appropriate feedback strategy, the previously proposed probabilistic scheme has been turned into a reiterative scheme which leads the target qubits to the singlet state with probability that converges asymptotically to $1$ in the number of iterations.
It is remarkable that no additional element or technology is required for the feedback: we have only to set the incident momentum of the ancilla off a resonance point.
The convergence to the unique fixed point (mixing property) is rigorously proved, and is shown to be robust again various types of imperfections in the scheme.
In particular, the scheme is very robust against the inefficiency of the detectors, which is clarified by a concise formula for the fidelity of the fixed point of the mixing map to the target singlet state.

We here concentrated on a specific physical model with two qubits fixed along a 1D channel, where ancilla qubits flow.
The present analysis however provides a general guideline for turning a probabilistic convergence scheme into a scheme that works with probability arbitrarily close $1$.
In general, the former probabilistic scheme keeps the target state, as long as the measurement on an ancilla reports desired result.
If the measurement outcome is not the desired one and if in such a case we are sure that the system is in an orthogonal state to the target, it is quite easy to design the feedback: we simply ``shake'' the system to provoke the transition from the orthogonal state to the target state.
This inflow to the target state, in the absence of the outflow, ensures the convergence to the target state.
Note also that the methods given in Ref.\ \cite{ref:DanielVittorio} can be interpreted as feedback schemes for Ref.\ \cite{ref:qpf}. 
The methods developed for these simple systems pave the way for general strategies for driving systems to target states.

\ack
This work is supported by a Special Coordination Fund for Promoting Science and Technology and the Grant-in-Aid for Young Scientists (B) (No.\ 21740294) both from the Ministry of Education, Culture, Sports, Science and Technology, Japan, 
by the Grant-in-Aid for Scientific Research (C) from the Japan Society for the Promotion of Science,
by the Italian Ministry of University and Research under the bilateral Italian-Japanese Projects II04C1AF4E on ``Quantum Information, Computation and Communication'' and the FIRB IDEAS project ESQUI, 
by the Joint Italian-Japanese Laboratory on ``Quantum Information and Computation'' of the Italian Ministry for Foreign Affairs,
and by the EPSRC grant EP/F043678/1.

\end{document}